\documentclass[aps,tightlines]{revtex4}%
\usepackage{amsfonts}
\usepackage{amsmath}
\usepackage{amssymb}
\usepackage{graphicx}%
\begin{document}
\title{Classification of Multipartite Entanglement via Negativity Fonts}
\author{S. Shelly Sharma}
\email{shelly@uel.br}
\affiliation{Departamento de F\'{\i}sica, Universidade Estadual de Londrina, Londrina
86051-990, PR Brazil }
\author{N. K. Sharma}
\email{nsharma@uel.br}
\affiliation{Departamento de Matem\'{a}tica, Universidade Estadual de Londrina, Londrina
86051-990 PR, Brazil }

\begin{abstract}
Partial transposition of state operator is a well known tool to detect quantum
correlations between two parts of a composite system. In this letter, the
global partial transpose (GPT) is linked to conceptually multipartite
underlying structures in a state - the negativity fonts. If $K-$way negativity
fonts with non zero determinants exist, then selective\ partial transposition
of a pure state, involving $K$ of the $N$ qubits ($K\leq N$) yields an
operator with negative eigevalues, identifying K-body correlations in the
state. Expansion of GPT interms of $K-$way partially transposed (KPT)
operators reveals the nature of intricate intrinsic correlations in the
state.{} Classification criteria for multipartite entangled states, based on
the underlying structure of global partial transpose of canonical state, are
proposed. Number of $N-$partite entanglement types for an $N$ qubit system is
found to be $2^{N-1}-N+2$, while the number of major entanglement classes is
$2^{N-1}-1$. Major classes for three and four qubit states are listed.
Subclasses are determined by the number and type of negativity fonts in
canonical state.

\end{abstract}
\maketitle

\section{Introduction}

Interactions generate correlated systems. Understanding the intricate nature
of correlations in multipartite systems is a fundamental problem of physics.
Correlations in multipartite systems that need quantum mechanics for their
description are conceptually distinct from classical correlations.
Entanglement is the best known aspect of such multipartite correlations.
Partial transposition of a state operator is a well known tool to detect
bi-partite entanglement. In this letter, we link the partial transpose to
conceptually multipartite underlying structures - the negativity fonts - and
present a new perspective to classification of multipartite entanglement.
Selective partial transposition involving a group of subsystems, allows one to
detect underlying intricate structure and nature of correlations through the
expansion of global partial transpose of the canonical state in terms of
$K$-way partially transposed operators. Each term in the expansion identifies
a specific type of multipartite correlations, which may in turn be quantified
by constructing appropriate invariantes.

The nature of quantum coherences generated by the interaction determines the
entanglement type of a multipartite state. Matrix elements of an N-partite
state operator that are off diagonal for $K$ subsystems in a selected basis,
determine the $K-$way coherences. Quantum coherences that are present in a
GHZ-like state of $K$ two-level (qubit) quantum systems may also be present in
an N qubit state, where $2\leq K\leq N$. We focus on classification of
multiqubit entanglement, although, analogous treatment should be valid for
subsystems with $d\neq2$. It was shown in \cite{shar07} that $K-$way
coherences in a multiqubit state can be quantified by partial $K-$way
negativities. Partial K-way negativities \cite{shar0809}, the contributions of
$K-$way coherences to global negativity \cite{zycz98,vida02}, vary under local
operations. How a $K-$way negativity font - the basic unit of $K-$qubit
coherences - transforms when one of the qubits undergoes local unitary
transformations was pointed out, recently, in \cite{shar101,shar102}. By using
local unitaries to express a general $N-$qubit state as a superposition of
minimum number of local basis product (LBP) states, one arrives at a canonical
state with characteristic number and type of negativity fonts quantifying
inherent $K-$way coherences $\left(  2\leq K\leq N\right)  $. The key idea is
to explore the inevitable link between the quantum coherences and entanglement
manifest in the global partial transpose (GPT) \cite{pere96}. All states with
the same type of intrinsic quantum coherences belong to the same class.
Furthermore, the number of negativity fonts with non-zero determinants, in a
$K-$way partially transposed canonical state, is fixed. Therefore, the states
in a given class can be grouped together into subclasses depending on the
number and type of distinct negativity fonts with non zero determinants in
global partially transposed canonical state.

Two multipartite pure states are considered equivalent under stochastic local
operations and classical communication (SLOCC), if one can be obtained from
the other with some probability using only local operations and classical
communication among different parties. Classification of three qubit states
focussed on local unitary invariance in ref. \cite{acin00}, and SLOCC
equivalence in \cite{dur00}.\ Classification of four qubit states into nine
families in \cite{vers02} and more recently from string theory approach in
\cite{bors10} has been expanded by Li et al. \cite{li09} to 49 SLOCC
entanglement classes. Lamata et al. \cite{lama07} use an inductive approach
\cite{lama06} to classify four-qubit pure states. Their main result is that
each of the eight genuine inequivalent entanglement classes contains a
continuous range of strictly nonequivalent states, although with similar
structure. It is clear that as we go to multipartite systems (or higher
dimensional systems), we are more likely to encounter continuous range of
strictly nonequivalent states, although with similar structure. In this
context, our classification criteria based on the nature of quantum
correlations, offer a distinct logical perspective. Local unitary invariants,
obtained from transformation equations for determinants of negativity fonts
under local unitaries \cite{shar102}, or SLOCC \cite{li07} quantify the
entanglement of states in a sub class. Classification of three and four qubit
states is reviewed.

\section{Basic concepts}

In order to clearly state our classification criteria, we introduce some
notations and definitions. A general $N-$qubit pure state reads as%
\begin{equation}
\left\vert \Psi^{A_{1},A_{2},...A_{N}}\right\rangle =\sum_{i_{1}i_{2}...i_{N}%
}a_{i_{1}i_{2}...i_{N}}\left\vert i_{1}i_{2}...i_{N}\right\rangle
\label{nqubit}%
\end{equation}
where $\left\vert i_{1}i_{2}...i_{N}\right\rangle $ is a local basis product
(LBP) state in $2^{N}$ dimensional Hilbert space, and $A_{p}$ is the location
of qubit $p$. The coefficients $a_{i_{1}i_{2}...i_{N}}$ are complex numbers.
The local basis states of a single qubit are labelled by $i_{m}=0$ and $1,$
where $m=1,...,N$. The global partial transpose\ $\widehat{\rho}_{G}%
^{T_{A_{p}}}$\cite{pere96}\textbf{\ }with respect to qubit $p$ is constructed
from $N$ qubit state operator $\widehat{\rho}$=$\left\vert \Psi\right\rangle
\left\langle \Psi\right\vert $ by partial transposition%
\begin{equation}
\left\langle i_{1}i_{2}...i_{N}\right\vert \widehat{\rho}_{G}^{T_{A_{p}}%
}\left\vert j_{1}j_{2}...j_{N}\right\rangle =\left\langle i_{1}i_{2}%
...i_{p-1}j_{p}i_{p+1}...i_{N}\right\vert \widehat{\rho}\left\vert j_{1}%
j_{2}...j_{p-1}i_{p}j_{p+1}...j_{N}\right\rangle .
\end{equation}
The $K-$way partial transpose, $\widehat{\rho}_{K}^{T_{A_{p}}}$, obtained by
selective transposition is defined as%
\begin{equation}
\left\langle i_{1}i_{2}...i_{N}\right\vert \widehat{\rho}_{K}^{T_{A_{p}}%
}\left\vert j_{1}j_{2}...j_{N}\right\rangle =\left\langle i_{1}i_{2}%
...i_{p-1}j_{p}i_{p+1}...i_{N}\right\vert \widehat{\rho}\left\vert j_{1}%
j_{2}...j_{p-1}i_{p}j_{p+1}...j_{N}\right\rangle , \label{ptk1}%
\end{equation}
if$\quad\sum\limits_{m=1}^{N}(1-\delta_{i_{m},j_{m}})=K,\quad$and$\quad
\delta_{i_{p},j_{p}}=0,$ while
\[
\left\langle i_{1}i_{2}...i_{N}\right\vert \widehat{\rho}_{K}^{T_{p}%
}\left\vert j_{1}j_{2}...j_{N}\right\rangle =\left\langle i_{1}i_{2}%
...i_{N}\right\vert \widehat{\rho}\left\vert j_{1}j_{2}...j_{N}\right\rangle
,
\]
if$\ \sum\limits_{m=1}^{N}(1-\delta_{i_{m},j_{m}})\neq K$. Decomposition of
global partial transpose, $\widehat{\rho}_{G}^{T_{A_{p}}},$ of an $N-$qubit
state (Eq. (\ref{nqubit})) with respect to qubit $A_{p}$ in terms of $K-$way
partially transposed operators reads as
\begin{equation}
\widehat{\rho}_{G}^{T_{A_{p}}}=\sum\limits_{K=2}^{N}\widehat{\rho}%
_{K}^{T_{A_{p}}}-(N-2)\widehat{\rho}. \label{3n}%
\end{equation}
Expansion of GPT interms of $K-$way partially transposed (KPT) operators has
information about the nature of intricate intrinsic correlations in the state.

\textit{Negativity fonts - }Following the notation used in ref. \cite{shar102}%
, an $N-$way negativity font and it's determinant read as
\[
\nu^{i_{1}i_{2}...i_{p}...i_{N}}=\left[
\begin{array}
[c]{cc}%
a_{i_{1}i_{2}...i_{p}...i_{N}} & a_{i_{1}+1,i_{2}+1,...i_{p}...i_{N}+1}\\
a_{i_{1}i_{2}...i_{p}+1...i_{N}} & a_{i_{1}+1,i_{2}+1,...i_{p}+1...i_{N}+1}%
\end{array}
\right]  ,D^{i_{1}i_{2}...i_{p}...i_{N}}=\det\left(  \nu^{i_{1}i_{2}%
...i_{p}...i_{N}}\right)  ,
\]
when the state of qubit $p$ is involved in partial transposition. Here
$i_{m}+1=0$ for $i_{m}=1$ and $i_{m}+1=1$ for $i_{m}=0$. When global partial
transpose is expressed as a sum of $K-$way partialy transposed operators, the
negativity fonts are distributed over $N-1$ operators. Since $K$ qubits may be
chosen in $\left(  \frac{N!}{\left(  N-K\right)  !K!}\right)  $ ways, the form
of a $K-$way font must specify the set of $K$ qubits it refers to. To
distinguish between different $K-$way negativity fonts a list of qubit states
for which $\delta_{i_{m},j_{m}}=1$ appears as a subscript of $\nu$. In other
words, a $K-$way font involving qubits $A_{q+1}$ to $A_{q+K}$ such that
$\sum\limits_{m=1}^{N}(1-\delta_{i_{m},j_{m}})=\sum\limits_{m=q+1}%
^{q+K}(1-\delta_{i_{m},j_{m}})=K$ reads as
\begin{align*}
&  \nu_{\left(  A_{1}\right)  _{i_{1}},\left(  A_{2}\right)  _{i_{2}%
},...\left(  A_{q}\right)  _{i_{q}}\left(  A_{q+K+1}\right)  _{i_{q+K+1}%
}...\left(  A_{N}\right)  _{i_{N}}}^{i_{1}i_{2}...i_{p}...i_{N}}\\
&  =\left[
\begin{array}
[c]{cc}%
a_{i_{1}i_{2}...i_{p}...i_{N}} & a_{i_{1}i_{2}...i_{q},i_{q+1}+1,i_{q+2}%
+1,...i_{p}...,i_{q+K}+1,i_{q+K+1},...i_{N}}\\
a_{i_{1}i_{2}...i_{p}+1...i_{N}} & a_{i_{1}i_{2}...i_{q},i_{q+1}%
+1,i_{q+2}+1,...i_{p}+1...,i_{q+K}+1,i_{q+K+1},...i_{N}}%
\end{array}
\right]  ,
\end{align*}
and its determinant is represented by $D_{\left(  A_{1}\right)  _{i_{1}%
},\left(  A_{2}\right)  _{i_{2}},...\left(  A_{q}\right)  _{i_{q}}\left(
A_{q+K+1}\right)  _{i_{q+K+1}}...\left(  A_{N}\right)  _{i_{N}}}%
^{i_{q+1}i_{q+2}...i_{p}...i_{q+k}}$. If $K-$way negativity fonts with non
zero determinants exist, then selective\ partial transposition of a pure
state, involving $K$ of the $N$ qubits ($K\leq N$) yields an operator with
negative eigevalues, identifying $K-$body correlations in the state. Local
unitaries on qubit $A_{r}$ with $r\neq p$, yield transformation equations
relating determinants of $K-$way and $\left(  K\pm1\right)  -$way negativity
fonts. Consequently, the contribution of partial $K-$way negativities to
global negativity varies with local unitaries.

\section{Canonical State}

A unitary transformation U$^{A_{1}}=\frac{\left\vert a_{11J}\right\vert
}{\sqrt{\left\vert a_{01J}\right\vert ^{2}+\left\vert a_{11J}\right\vert ^{2}%
}}\left[
\begin{array}
[c]{cc}%
1 & -\frac{a_{01J}}{a_{11J}}\\
\frac{a_{01J}^{\ast}}{a_{11J}^{\ast}} & 1
\end{array}
\right]  $, followed by an invertible local operator transforms an N-way
negativity font in $\rho_{N}^{T_{A_{1}}}$ as
\[
\left[
\begin{array}
[c]{cc}%
a_{00I} & a_{01J}\\
a_{10I} & a_{11J}%
\end{array}
\right]  \underrightarrow{U^{A_{1}}}\left[
\begin{array}
[c]{cc}%
b_{00I} & 0\\
b_{10I} & b_{11J}%
\end{array}
\right]  \underrightarrow{O^{A_{1}}}\left[
\begin{array}
[c]{cc}%
1 & 0\\
0 & c_{11J}%
\end{array}
\right]
\]
such that $b_{00I}b_{11J}=a_{00I}a_{11J}-a_{10I}a_{01J}$ and entanglement type
of the state is not changed. Here, $I$ stands for the string $\left\{
i_{3}i_{4}...i_{N}\right\}  $. Therefore, the most general $N-$qubit state
with $2^{N-2}$ $N-$way fonts can be transformed\ by LOCC to a unitary
equivalent canonical form which is a superposition of $2^{N}-N$ LBP States.
The state having information about the entanglement type has at the most
$2^{N}-2N$ LBP States.

A unitary transformation U$^{A_{q}}$ (applied to a qubit other than $A_{1}$)
may be selected such that
\[
\det\left[
\begin{array}
[c]{cc}%
a_{00I} & a_{01J}\\
a_{10I} & a_{11J}%
\end{array}
\right]  \underrightarrow{U^{A_{q}}}\det\left[
\begin{array}
[c]{cc}%
c_{00I} & c_{01J}\\
c_{10I} & c_{11J}%
\end{array}
\right]  =0
\]
thus annihilating a negativity font. In an entangled state, the maximum number
of negativity fonts that may be annihilated by unitary operations and
classical communication is $N-1$. The final canonical state obtained after
local invertible operations and classical communication will have a variable
number of negativity fonts and LBP states. For states with $N>3$ more than one
representations of canonical state may be possible.

\section{Criterion for classifying N-qubit entangled States}

A logical criterion for classification of multipartite entanglement can be
based on the structure of global partial transpose (Eq. (\ref{3n})) of
canonical state and number of $K-$way negativity fonts. Determinants of
negativity fonts are the intrinsic negative eigenvalues of partially
transposed state operator. No wonder, squared global negativity is four times
the sum of squared moduli of determinants of all negativity fonts
\cite{shar11} available for a given qubit. Combinations of Determinants of
Negativity fonts determine the local unitary invariants characterizing an
N-qubit state \cite{shar101,shar102}. These observations lead us to a
classification scheme in which

(a) An entanglement class is characterized by the set of $K-$way $\left(
2\leq K\leq N\right)  $\ partially transposed operators present in the
expansion of global partial transpose of the canonical state $\widehat{\rho
}_{c}$. Global partial transpose of canonical state has the decomposition
$\left(  \widehat{\rho}_{c}\right)  _{G}^{T_{A_{p}}}=%
{\textstyle\sum\limits_{K=2}^{N}}
\left(  \widehat{\rho}_{c}\right)  _{K}^{T_{A_{p}}}-(N-2)\widehat{\rho}_{c}$.
Since $K$ varies from $2$ to $N,$ the number of possible combinations of
$K-$way partially transposed matrices is
\[
N_{Class}=%
{\textstyle\sum\limits_{m=1}^{N-1}}
\frac{\left(  N-1\right)  !}{m!\left(  N-m-1\right)  !}=\left(  2^{N-1}%
-1\right)  ,
\]
which is also the number of major entanglement classes. An $N-$partite state
is said to be $N-$partite entangled if and only if all bipartite partitions
produce mixed reduced density matrices. Besides the fully separable and fully
entangled, there also exists classes of states that are partially separable.
The case $\widehat{\rho}_{G}^{T_{A_{p}}}=\widehat{\rho}_{K}^{T_{A_{p}}}$
results in an $N-$partite entangled state, only if $K=N$ or $K=2$. Therefore
the number of $N-$partite entanglement types is $\left(  2^{N-1}-N+2\right)
$. States obtained by permuting the qubits are grouped in the same class.

(b) States with similar decomposition of $\left(  \widehat{\rho}_{c}\right)
_{G}^{T_{A_{p}}}$ with respect to different number of qubits belong to
different sub classes or families in the same class. For a given qubit, the
number of $K-$way negativity fonts in a $K-$way partially transposed matrix
varies from $0$ to $2^{N-2}$. The number and types of non-zero determinants of
distinct negativity fonts in $KPT$ canonical state operators characterize
nonequivalent sub classes of states.

(c) The values of entanglement monotones based on relevant local unitary
polynomial invariants quantify the entanglement of a state in a sub-class.

\textit{Special Entanglement types of N--qubit States - }From criterion (a),
it is natural to find states with similar entanglement type but different
number of qubits. For instance, all $N$- qubit GHZ like states are
characterized by $\left(  \widehat{\rho}_{c}\right)  _{G}^{T_{A_{p}}}=\left(
\widehat{\rho}_{c}\right)  _{N}^{T_{A_{p}}}$, while for $N$ qubit W-like
states $\left(  \widehat{\rho}_{c}\right)  _{G}^{T_{A_{p}}}=\left(
\widehat{\rho}_{c}\right)  _{2}^{T_{A_{p}}}$. Other entanglement types can be
identified by using the transformation properties of determinants of
negativity fonts and unitary equivalence. For example, a common feature of a
special set of states with $\left(  \widehat{\rho}_{c}\right)  _{G}^{T_{A_{p}%
}}=\left(  \widehat{\rho}_{c}\right)  _{N}^{T_{A_{p}}}+\left(  \widehat{\rho
}_{c}\right)  _{2}^{T_{A_{p}}}-\widehat{\rho}_{c}$ is a unitary equivalent
description with $\left(  \widehat{\rho}_{c}\right)  _{G}^{T_{A_{p}}}=\left(
\widehat{\rho}_{c}\right)  _{N-1}^{T_{A_{p}}}+\left(  \widehat{\rho}%
_{c}\right)  _{2}^{T_{A_{p}}}-\widehat{\rho}_{c}$. To illustrate, we consider,
the global partial transpose of an $N-$qubit state with canonical form%
\begin{equation}
\left\vert \Psi\right\rangle =a_{00...0}\left\vert 00...0\right\rangle
+a_{11...1}\left\vert 11...1\right\rangle +a_{0011...1}\left\vert
0011...1\right\rangle +a_{1100...0}\left\vert 1100...0\right\rangle ,
\label{cluster}%
\end{equation}
which is characterized by $N-$way, $\left(  N-2\right)  $-way and $2-$way
negativity fonts. The combination $\left(  D^{00...0}-D^{0011...1}\right)
^{2}-4D_{\left(  A_{2}\right)  _{0}}^{00...0}D_{\left(  A_{2}\right)  _{1}%
}^{00...0}$ is invariant with respect to local unitaries on qubits A$_{1}$ and
A$_{2}.$\ If the determinants of two $N-$way fonts satisfy, $D^{00...0}%
+D^{0011...1}=0,$then the form of invariant indicates the existence of local
unitaries on A$_{1}$ and A$_{2}$ that transform the state to a form for which
GPT has $\left(  N-1\right)  -$way negativity fonts. For the state of Eq.
(\ref{cluster}) the unitary equivalent form reads as%
\[
\left\vert \Psi^{\prime}\right\rangle =b_{00...0}\left\vert
00...0\right\rangle +b_{1011...1}\left\vert 1011...1\right\rangle
+b_{0111...1}\left\vert 0111...1\right\rangle +b_{1100...0}\left\vert
1100...0\right\rangle .
\]
Four qubit cluster state $\left\vert C\right\rangle =\left\vert
0000\right\rangle -\left\vert 0011\right\rangle +\left\vert 1110\right\rangle
+\left\vert 1101\right\rangle ,$belongs to the class with $\left(
\widehat{\rho}_{c}\right)  _{G}^{T_{A_{p}}}=\left(  \widehat{\rho}_{c}\right)
_{4}^{T_{A_{p}}}+\left(  \widehat{\rho}_{c}\right)  _{2}^{T_{A_{p}}}%
-\widehat{\rho}$. In general, when two equivalent canonical descriptions are
possible the one with higher order negativity fonts should determine the class.

\begin{table}[ptb]
\caption{Classification of three qubit States}%
\label{t1}
\begin{tabular}
[c]{||c||c||c||c||c||}\hline\hline
Class & Decomposition of $\left(  \rho_{c}\right)  _{G}^{T_{A_{p}}}$ &
Representative $\rho_{c}$ & $\left\{  N_{3-way},N_{2-way}\right\}  $ &
3-tangle\\\hline\hline
CI & $\left(  \rho_{c}\right)  _{3}^{T_{A_{p}}}+\left(  \rho_{c}\right)
_{2}^{T_{A_{p}}}-\widehat{\rho}_{c}$ & $\left\vert 000\right\rangle
+\left\vert 111\right\rangle +\left\vert 110\right\rangle $ & $\left\{
1,1\right\}  $,$\left\{  1,2\right\}  $,$\left\{  1,3\right\}  $ & $\tau
_{3}\neq\left(  N_{G}^{A_{p}}\right)  ^{2}$\\\hline\hline
CII & $\left(  \rho_{c}\right)  _{3}^{T_{A_{p}}}$ & $\left\vert
000\right\rangle +\left\vert 111\right\rangle $ & $\left\{  1,0\right\}  $ &
$\tau_{3}=\left(  N_{G}^{A_{p}}\right)  ^{2}$\\\hline\hline
CIII & $\left(  \rho_{c}\right)  _{2}^{T_{A_{p}}}$ & $\left\vert
000\right\rangle +\left\vert 110\right\rangle +\left\vert 101\right\rangle $ &
$\left\{  0,1\right\}  $,$\left\{  0,2\right\}  $ & $\tau_{3}=0$\\\hline\hline
\end{tabular}
\end{table}

\section{Counting the Classes and Sub classes}

For the most general three qubit state, local unitaries yield the canonical
form\cite{acin00}%
\begin{equation}
\left\vert \Psi_{c}\right\rangle =a_{000}\left\vert 000\right\rangle
+a_{100}\left\vert 100\right\rangle +a_{111}\left\vert 111\right\rangle
+a_{101}\left\vert 101\right\rangle +a_{110}\left\vert 110\right\rangle
,\label{3can}%
\end{equation}
having a single $3-$way negativity font in $\widehat{\rho}_{G}^{T_{A_{1}}}$
and five LBPS. Global partial transpose of $\widehat{\rho}=\left\vert \Psi
_{c}\right\rangle \left\langle \Psi_{c}\right\vert $ has the decomposition
$\left(  \widehat{\rho}_{c}\right)  _{G}^{T_{p}}=\left(  \widehat{\rho}%
_{c}\right)  _{3}^{T_{p}}+\left(  \widehat{\rho}_{c}\right)  _{2}^{T_{p}%
}-\left(  \widehat{\rho}_{c}\right)  $. Table I displays the three major
classes, distinguished by the relation between%
\[
\tau_{3}=4\left\vert \left(  \left(  D^{000}+D^{001}\right)  ^{2}-4D_{\left(
A_{3}\right)  _{0}}^{00}D_{\left(  A_{3}\right)  _{1}}^{00}\right)
\right\vert ,
\]
and global negativity of GPT, along with $\left\{  N_{3-way},N_{2-way}%
\right\}  $ that characterize the sub-classes of three qubit entangled states.
Here $N_{K-way}$ is the number of $K-$way negativity fonts with non-zero determinants.

Applying the classification criterion (a) to four qubit states we obtain seven
major classes of states of which six contain states with four qubits entangled
to each other. Table II. displays a representative of each class and possible
number of $\left\{  N_{4-way},N_{3-way}\right\}  $. Sub classes in a major
class have different number of LBP states and characteristic combination of
$\left\{  N_{4-way},N_{3-way},N_{2-way}\right\}  $ in canonical state. To
distinguish between the states in the same sub-class, one resorts to four
qubit invariants. We notice that states with four-partite entanglement arise
in six of the seven classes that is there are six distinct ways of entangling
four qubits. In comparison, Lamata claim eight distinct entanglement types and
Acin et al. \cite{acin00} claim nine distinct entanglement types. The
observation of Lamata et al. \cite{lama07} that the class $L_{0_{3\oplus
1}0_{3\oplus1}}$ \cite{acin00} with canonical state $\left\vert
0000\right\rangle +\left\vert 1110\right\rangle $ does not contain genuinely
entangled four qubit states is consistent with class VI\ in table II. In
addition, we notice that the states with canonical forms $\left\vert
0000\right\rangle +\left\vert 1111\right\rangle +\left\vert 1100\right\rangle
$ and $\left\vert 0000\right\rangle +\left\vert 1111\right\rangle +\lambda
_{1}\left\vert 1100\right\rangle +\lambda_{2}\left\vert 0011\right\rangle $
differ only in the number of four way negativity fonts, while both have a
canonical form with $\left(  \widehat{\rho}_{c}\right)  _{G}^{T_{p}}=\left(
\widehat{\rho}_{c}\right)  _{4}^{T_{p}}+\left(  \widehat{\rho}_{c}\right)
_{2}^{T_{p}}-\left(  \widehat{\rho}_{c}\right)  $. Like wise the states with
$\left(  \widehat{\rho}_{c}\right)  _{G}^{T_{p}}=\left(  \widehat{\rho}%
_{c}\right)  _{4}^{T_{p}}+\left(  \widehat{\rho}_{c}\right)  _{3}^{T_{p}%
}+\left(  \widehat{\rho}_{c}\right)  _{2}^{T_{p}}-2\left(  \widehat{\rho}%
_{c}\right)  $, but different number of negativity fonts belong to different
sub-classes in our classification, whereas are classified as different
entanglement types by Lamata et al. \cite{lama07} (span $\left\{  0_{k}%
\Psi,GHZ\right\}  $, some of the states from span $\left\{  0_{k}\Psi
,0_{k}\Psi\right\}  $ and span $\left\{  GHZ\text{, }W\right\}  $).
\begin{table}[ptb]
\caption{Seven Classes of Four qubit States.}%
\label{t2}
\begin{tabular}
[c]{||c||p{1.2in}||c||c||}\hline\hline
Class & \multicolumn{1}{||c||}{Decomposition of $\left(  \rho_{c}\right)
_{G}^{T_{A_{p}}}$} & Representative $\widehat{\rho}_{c}$ & $\left\{
N_{4-way},N_{3-way}\right\}  $\\\hline\hline
CI & $\left(  \rho_{c}\right)  _{4}^{T_{A_{p}}}+\left(  \rho_{c}\right)
_{3}^{T_{A_{p}}}+\left(  \rho_{c}\right)  _{2}^{T_{A_{p}}}-2\widehat{\rho}%
_{c}$ & $\left\vert 0000\right\rangle +\left\vert 1111\right\rangle
+\left\vert 1110\right\rangle +\left\vert 1100\right\rangle $ & $\left\{
\left(  3-1\right)  ,\left(  1-12\right)  \right\}  $\\\hline\hline
CII & \multicolumn{1}{||c||}{$\left(  \rho_{c}\right)  _{4}^{T_{A_{p}}%
}+\left(  \rho_{c}\right)  _{3}^{T_{A_{p}}}-\widehat{\rho}_{c}$} & $\left\vert
0000\right\rangle +\left\vert 1111\right\rangle +\left\vert 1110\right\rangle
$ & $\left\{  1,1\right\}  $\\\hline\hline
CIII & \multicolumn{1}{||c||}{$\left(  \rho_{c}\right)  _{4}^{T_{A_{p}}%
}+\left(  \rho_{c}\right)  _{2}^{T_{A_{p}}}-\widehat{\rho}_{c}$} & $\left\vert
0000\right\rangle +\left\vert 1111\right\rangle +\left\vert 0011\right\rangle
-\left\vert 0011\right\rangle $ & $\left\{  \left(  4-2\right)  ,0\right\}
$\\\hline\hline
CIV & \multicolumn{1}{||c||}{$\left(  \rho_{c}\right)  _{3}^{T_{A_{p}}%
}+\left(  \rho_{c}\right)  _{2}^{T_{A_{p}}}-\widehat{\rho}_{c}$} & $\left\vert
0000\right\rangle +\left\vert 1110\right\rangle +\left\vert 1101\right\rangle
$ & $\left\{  0,1-4\right\}  $\\\hline\hline
CV & \multicolumn{1}{||c||}{$\left(  \rho_{c}\right)  _{4}^{T_{A_{p}}}$} &
$\left\vert 0000\right\rangle +\left\vert 1111\right\rangle $ & $\left\{
1,0\right\}  $\\\hline\hline
CVI & \multicolumn{1}{||c||}{$\left(  \rho_{c}\right)  _{3}^{T_{A_{p}}}$} &
$\left\vert 0000\right\rangle +\left\vert 1110\right\rangle ;($only
Separable$)$ & $\left\{  0,1\right\}  $\\\hline\hline
CVII & \multicolumn{1}{||c||}{$\left(  \rho_{c}\right)  _{2}^{T_{A_{p}}}$} &
$\left\vert 0000\right\rangle +\left\vert 1100\right\rangle +\left\vert
1010\right\rangle +\left\vert 1001\right\rangle $ & $\left\{  0,0\right\}
$\\\hline\hline
\end{tabular}
\end{table}For five qubits States one obtains fifteen classes of which 13
contain states with all five qubits entangled to each other.

\section{Conclusions}

Using the structure of global partial transpose of the canonical state as a
key to the nature of intrinsic quantum correlations, we have proposed criteria
for classifying $N$ qubit pure states. The set of criteria offers a new
perspective to entanglement classification and is extendible to mixed state
entanglement. It has been pointed out that for different number of qubits one
may identify states with similar quantum correlations. Most of the major
classes contain a continuous range of strictly SLOCC nonequivalent states with
similar structure. The number of genuine entanglement types for a given value
of N is easily counted. As the number of $N-$way negativity fonts increases,
so does the number of sub-classes. However, the distinct number of new
entanglement types does not grow so fast. It will be interesting to study the
new features, such as suitability for a given QIP task, that appear with
increasing $N$ for states with same type of intrinsic quantum correlations.

This work is supported by CNPq Brazil and FAEP UEL Brazil.

\end{document}